\documentclass[]{elsart3p}

\usepackage{epsfig}

\usepackage{amsfonts}
\usepackage{amssymb}

\usepackage{lineno}

\newcommand{\isotope}[2][]{\ensuremath{^{\mathrm{#1}}\mathrm{#2}}}

\begin{document}

\begin{frontmatter}



\title{A \isotope[13]{C}($\alpha$,\isotope{n})\isotope[16]{O}
  calibration source for KamLAND}


\author{David W. McKee}\ead{dmckee@bama.ua.edu}, 
\author{Jerome K. Busenitz\corauthref{cor1}}\ead{busenitz@bama.ua.edu},
\author{Igor Ostrovskiy}\ead{iostrovskiy@bama.ua.edu}
\corauth[cor1]{Department of Physics and Astronomy; University of
Alabama; Box 870324; Tuscaloosa, AL 35487}

\address{University of Alabama}

\begin{abstract}
  We report on the construction and performance of a calibration
  source for KamLAND using the reaction
  \isotope[13]{C}($\alpha,\isotope{n})\isotope[16]{O}$ with
  \isotope[210]Po as the alpha progenitor.  The source provides a
  direct measurement of this background reaction in our detector,
  high energy calibration points for the detector energy scale,
  and data on quenching of the neutron visible energy in KamLAND
  scintillator. We also discuss the possibility of using the
  $\isotope[13]{C}(\alpha,\isotope{n})\isotope[16]{O}$ reaction
  as a source of tagged slow neutrons.
\end{abstract}

\end{frontmatter}

\section{Introduction}
\label{intro}
The reactor phase of the Kamioka Liquid scintillator
Anti-Neutrino Detector (KamLAND) detects the interaction
\begin{equation}
\bar{\nu} + \isotope{p} \rightarrow \isotope{n} + \isotope{e}^{+}
\end{equation} 
by observing the delayed coincidence between the positron and the
subsequent capture of the
neutron.\cite{Araki:2004mb} A trigger can be
generated by any reaction which produces a prompt signal above
threshold and a free neutron.

A particular background in KamLAND results from the presence of
\isotope[210]{Po} (a daughter of \isotope[222]{Rn} present in
quantities governed by the \isotope[210]{Pb} concentration),
which decays by emitting a 5.304~MeV $\alpha$ particle, allowing
the alpha capture reaction
\begin{equation}
  \label{alphaCapture}
\alpha + \isotope[13]{C} \rightarrow n + \isotope[16]{O}.
\end{equation}
This reaction has been studied since the 1950s,
\cite{Walton:1957,Sekharan:1967,Bair:1973su,Jacobs:1983} and
Harrisopulos et~al. recently measured the cross-section to
4\%.\cite{Harissopulos:2005cp} However, the current theoretical
understanding of the reaction was insufficient to describe the
response of KamLAND to the reaction in Eq.~\ref{alphaCapture} due
to uncertainties in the neutron spectrum and the poorly known
excited state branching fractions.

We have produced and deployed a calibration source utilizing the
reaction in Eq.~\ref{alphaCapture} to carry out a direct
measurement of the background rate and the energy spectrum for
this reaction in KamLAND.  Similar sources reported in the extant
literature\cite{Dickens:1970,Geiger:1978,Mason:1985} use
different progenitor isotopes and were constructed as gamma
calibration sources for germanium detectors; while we are as
interested in the neutron as the decay gamma.

\section{Design and Construction}
\label{design}
To obtain an accurate measurement of the prompt neutron energy
spectrum in KamLAND requires either \isotope[210]{Pb} or
\isotope[210]{Po} as the progenitor---other isotopes generate
different alpha energies.  Despite the short half-life of the
Polonium isotope---138.4 days---we judged it feasible to
construct a source, certify it for use in the detector, and
deploy it on a time scale comparable with the \isotope[210]{Po}
half-life. After our attempts to obtain \isotope[210]{Pb} in
sufficient quantities were unsuccessful, we decided to build a
source with \isotope[210]{Po}.

The strongest design constraints were imposed by the stringent local
regulatory limits on contained activity and the very low capture
fraction for alpha particles. We were limited to an initial contained
activity of 100~$\mu$Ci of \isotope[210]{Po}. Assuming this limit and
computing the expected capture fraction (see
section~\ref{captureFraction}) we could estimate a neutron rate for
the source no higher than 30 Hz.

We were able to obtain \isotope[210]{Po} in a 4~M hydrochloric
acid solution and high purity \isotope[13]{C} in powder form. The
source was constructed by filling the capsule with approximatly
0.3~g of \isotope[13]{C} powder, dripping the Polonium solution
into the carbon powder, and allowing the whole to dry thoroughly
before tamping the powder with a Delrin spacer and closing the
system. A heat lamp was used to speed the evaporation, which
required two days. Our design called for a total contained
activity of 95 $\mu$Ci on the day of assembly.

\begin{figure}[h]
  \centering
  \resizebox{\columnwidth}{!}{%
    \includegraphics{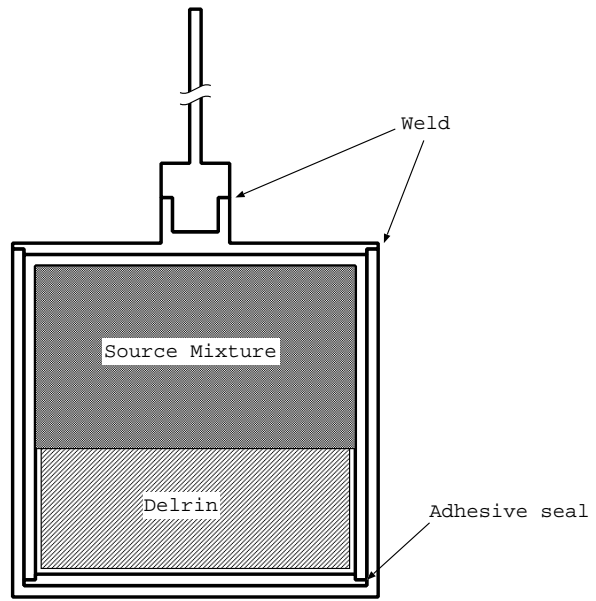}%
  }
  \caption{
    \label{capsulefig}
    Construction of the source capsule. The chamber is 13~mm in
    diameter and 13~mm in height. Both capsules have 1~mm thick
    walls. The inner capsule was inserted with the lid away from
    the joint in the outer capsule to protect the adhesive from
    the heat of welding.  }
\end{figure}

The source capsule---shown in Fig.~\ref{capsulefig}---was
constructed of stainless steel for ease of manufacture. The
inner capsule was constructed of series 316L stainless steel for
acid resistance, and series 304 stainless steel was used in the
outer capsule for its good welding properties. Both materials are
known to be compatible with KamLAND liquid scintillator (LS).

Despite the use of low carbon stainless steel in the inner capsule,
tests showed that it would nonetheless develop gas bubbles in the
presence of HCl that threatened to spatter our alpha source around the
fume hood. To prevent this, we painted the inside of the capsule with
four thin coats of a clear acrylic paint obtained at a local craft
store. This treatment provided adequate acid resistance.

The inner capsule was sealed using a structural
adhesive,\footnote{3M Scotch Grip 1357} wiped clean, and inserted
in the outer capsule, which was welded shut using an electron
beam technique.

\section{Certification}
\label{certification}
Objects to be deployed in KamLAND must first be certified as both
chemically compatible with the LS and radiologically clean. In
particular, it is necessary to show that radiological sources are
properly sealed and do not leak. 

Following standard KamLAND procedure, we thoroughly cleaned the
source, and then soaked it in 0.1~M nitric acid for four days,
pressure cycling to five atmospheres three times in the course of
the soak. The soak liquid was counted in a high sensitivity
germanium detector to exclude gamma radio-contamination. Special
attention was paid to the possible presence of \isotope[40]{K},
and the daughters of \isotope[238]{U} and \isotope[232]{Th}.

Because \isotope[210]Po does not have a significant gamma line, this
method is not well suited to detecting a low activity Polonium
leak. Instead we introduced a sample of the soak liquid into a cuvet
full of acid tolerant liquid scintillator,\footnote{Packard Ultima
  Gold AB} which was subsequently placed between two PMTs and the
signal from any alpha activity observed directly. Understanding this
device required calibration data with several gamma sources to
establish the energy response; with a clean control sample to
understand the shape of the background (mostly cosmic rays and ambient
radioactivity); and with a \isotope[210]Po doped sample to establish
the quenching behavior of the LS. A null result was obtained for
measured \isotope[210]{Po} leakage with a 90\%~C.L. upper limit of
0.3~Bq, and the source was certified for use in KamLAND.

\section{Source Physics}
\label{physics}
KamLAND signals from the source arise from three mechanisms:
prompt activity from the alpha progenitor, prompt activity from
the alpha capture reaction, and delayed activity from the capture
of the thermalized neutron.

At the alpha energy of \isotope[210]{Po}, Eq.~\ref{alphaCapture}
can proceed not only to the ground state of \isotope[16]{O}, but
also to the first two excited states. See Table~\ref{o16states}
for thresholds and decay products.

\subsection{Progenitor activity}
The \isotope[210]{Po} progenitor decays primarily by
\begin{eqnarray}
\isotope[210]{Po} \rightarrow \isotope[206]{Pb} & + & \alpha (5.304\mbox{
MeV}), \nonumber
\end{eqnarray}
but has a $(1.21\pm0.04) \times 10^{-5}$ branch to
\begin{eqnarray} 
  \label{decayGamma}
  \isotope[210]{Po} \rightarrow \isotope[206]{Pb} & + & 
  \alpha (4.517\mbox{ MeV}) \nonumber\\
   & + & \gamma (0.803\mbox{ MeV}). 
\end{eqnarray} 
The gamma line proved to be useful for calibrating the total
contained activity of the source. See section~\ref{activity}.

\subsection{Prompt signal}
The final state of Eq.~\ref{alphaCapture} contains two particles,
either of which can contribute to a prompt detector signal. In
practice the visible energy is mainly due to one particle. Captures to
the ground state of \isotope[16]{O} result in energetic neutrons which
will often generate enough scintillation light to pass the prompt
trigger threshold. The neutrons associated with excited state captures
are too low in energy to generate a trigger in this way, but the decay
products of the excited oxygen nuclei are well above threshold.

The neutrons emitted from captures to the ground state have a
broad energy spectrum between approximately 3.0 and 7.3~MeV,
peaking near 4.8~MeV.  The scattering of fast neutrons off of
protons in the liquid scintillator can produce a PMT signal
sufficient to develop a prompt trigger without contribution from
other mechanisms. Furthermore, a fast neutron can lose a large
fraction of its energy in a single inelastic scattering event on
carbon by
$$
n + \isotope[12]{C} \rightarrow n + \isotope[12]{C}^{*}
$$ 
followed by a prompt decay of the excited carbon nucleus with
emission of a 4.4~MeV $\gamma$.

\subsection{Delayed signal}
As in the neutrino events KamLAND was designed to detect, thermalized
neutrons are captured primarily by \isotope[1]{H} or rarely by
\isotope[12]{C}. These processes produce signals of 2.2 and 4.95~MeV
respectively. The mean neutron capture time in the detector is
slightly more than $200\mbox{ }\mu\mbox{s}$.


\section{Deployment and Data Set}
\label{data}
Once the source was certified for use in KamLAND, we deployed it
to a number of points along the vertical symmetry axis of the
detector in November 2006 and again in March of 2007. The source
was also deployed off the symmetry axis during January
2007. Though some 16 positions along the vertical axis and
several off axis geometries were probed, this analysis concerns
itself with runs taken at the detector center.

KamLAND triggers are based on the number, NSUM, of PMTs detecting
at least one photo-electron and the data acquisition (DAQ)
hardware can limit the total DAQ rates by dividing each second
into active and inactive windows. We had two separate needs: to
accurately measure the contained activity of the device (using
the decay gamma line in Eq.~\ref{decayGamma}) and to capture as
much data on the prompt and delayed spectra as possible. For the
first need we used a simple trigger---denoted the `activity'
trigger---which captured 35\% of all events with $\mbox{NSUM} >
70$ (i.e. with gamma equivalent energy of $\approx 250\mbox{
  keV}$ at the detector center). For the second goal we employ
the so-called `spectrum' trigger which captures all events with
$\mbox{NSUM} > 190$ and 1\% of all other events down to NSUM of
40.

\section{Source Performance Analysis}
\label{analysis}
The data from the polonium--carbon source are being used by the
KamLAND collaboration to characterize the
$\isotope[13]C(\alpha,\isotope{n})\isotope[16]{O}$ background to
the experiment, to better understand the properties of the
KamLAND liquid scintillator, and to tune the performance of our
event reconstruction algorithms. Here we present results of an
analysis to measure the rates from the source in order to
characterize it and demonstrate that it performs as expected.

All analysis includes a geometric cut constraining our
consideration to events reconstructed within 120~(150)~cm of the
source position for prompt events (neutron captures) and a timing
cut excluding the 2~ms following the detection of a muon (i.e. a
cosmic ray event which swamps the DAQ). These cuts reduce the
total data set considerably but retain all source-related events.

\begin{figure}[h]
  \centering
  \resizebox{\columnwidth}{!}{%
    \includegraphics{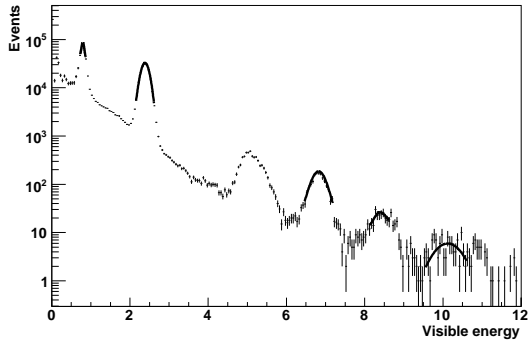}%
  }
  \caption{
    \label{singlesfig}
    Reconstructed singles energy spectrum. The spectrum has been
    corrected to allow for dead-time below NSUM of 190. The fits
    are (left to right) the 803~keV gamma, the neutron capture on
    hydrogen, the 6.13~MeV gamma from the second excited state of
    \isotope[16]{O}, and two background complexes from neutron
    capture on \isotope{Fe}, \isotope{Ni}, and \isotope{Co}. The
    unfit peak at 5~MeV includes both neutron capture on carbon
    and neutron inelastic scattering on carbon. The 1022~keV peak
    is too small to be visible on this plot. The continuum is
    largely the response of the LS to fast neutrons. The energy
    scale of the plot is linear in collected charge and
    calibrated to the 2.2~MeV neutron capture peak. 
  }
\end{figure}

Figure~\ref{singlesfig} displays a singles spectrum after the
geometric and muon cuts have been applied. Gaussians have been
fit to several of the interesting features of the spectrum.

\subsection{Contained \isotope[210]Po Activity}
\label{activity}
Measurement of the contained activity consists of fitting a
Gaussian plus linear background to the reconstructed energy near
803 keV in runs using the activity trigger.

Monte~Carlo studies indicate that $15\pm1$\% of events suffer
significant scattering losses in the source capsule and are not
fit.

\begin{figure}[h]
  \centering
  \resizebox{\columnwidth}{!}{%
    \includegraphics{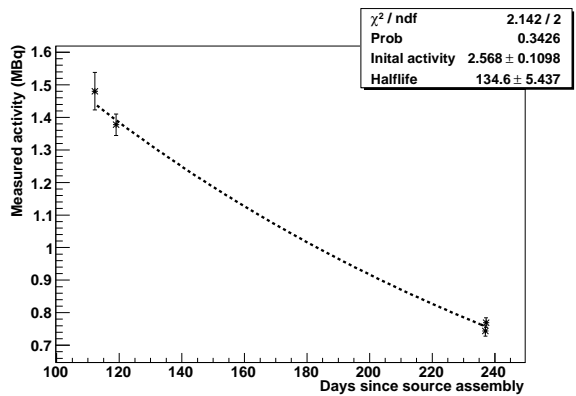}%
  }
  \caption{
    \label{decayfig}
    Contained \isotope[210]{Po} activity of the source as
    measured in KamLAND. Activity is computed from the area of a
    Gaussian fit to the peak at 803~keV, and is not corrected for
    scattering losses in the source capsule.}
\end{figure}

The data are sufficient to fit an exponential decay curve as shown in
Fig.~\ref{decayfig}. The fit yields a half-life for the source of
$135\pm5$ days---consistent with the accepted value for
\isotope[210]{Po}---and gives the initial contained activity as
$3.02\pm0.13\pm0.10\mbox{ MBq}= 81.6\pm3.5\pm2.7\mbox{ }\mu\mbox{Ci}$
of \isotope[210]{Po} correcting for the fitting fraction (the first
error represents the fitting error for the 803~keV peak including
statistics; the second represents the uncertainties of the branching
ratio and the fitting losses taken in quadrature). This is in
reasonable agreement with the design activity of $95\mbox{
}\mu\mbox{Ci}$ taking into account that some loss of activity by
adherence to the glassware during fabrication could be expected.

\subsection{Delayed Coincidence Rate}
\label{coincidence}
\begin{figure}[h]
  \centering
  \resizebox{\columnwidth}{!}{%
    \includegraphics{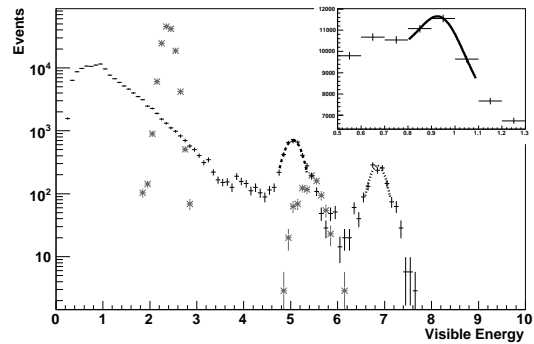}%
  }
  \caption{
    \label{decompfig}
    Reconstructed coincidence energy spectra.  Plotted are both
    the prompt (black points) and delayed (gray asterisks) energy
    spectra. The insert shows an expanded view of the prompt
    spectrum in the region of the 1022~keV line. The lines are
    fits to the positron annihilation peak (solid), the carbon
    inelastic peak (dashed), and the 6.13~MeV gamma events from
    the second excited state (dotted).
    The coincidence selection algorithm applies cuts on the
    delayed event energy, so no continuum is present between the
    hydrogen and carbon capture peaks.
  }
\end{figure}

We find prompt--delayed coincidences by looking for two events
separated by less than $1.5\mbox{ ms}$ and having a reconstructed
energy in the later event near 2.2 or 4.95 MeV. In practice, we
identify candidate neutron captures first, then work backwards in
time searching for corresponding prompt events. A example of the
results appears in Fig.~\ref{decompfig}.

This method is used to select pairs for all subsequent
analysis. Further, we associate a weight with each event to correct
for the trigger fraction. Monte~Carlo simulations of the trigger
behavior are used to establish the correct weight for each trigger and
NSUM range. We measured delayed coincidence rates of approximately
10~Hz in November 2006 and 5.5~Hz in April 2007. These results do not
include the neutrons captured on \isotope{Fe}, \isotope{Co},
\isotope{Ni} in the source capsule and the deployment hardware
(totaling a few percent of the rate) nor does it account for the
selection efficiency of the analysis ($\approx 1.5\%$). The measured
coincidence rates are consistent with the measured \isotope[210]{Po}
activity.

\subsection{Alpha Capture Fraction}
\label{captureFraction}
Computing the delayed coincidence rate as a function of the contained
activity averaged across all runs we find $(6.1\pm0.3)\mbox{
  events/s/(MBq of \isotope[210]{Po})}$. The expected alpha capture
fraction from integrating the cross-sections and stopping powers
reported in the literature is $(7.1\pm0.3)\mbox{ events/s/(MBq of
  \isotope[210]{Po})}$, neglecting any edge effects in the source
capsule and assuming that there has been no formation of \isotope{Po}
slugs in the source mixture. We compute the neutron activity on the
day of assembly to have been $18.4\pm1.3\mbox{ Bq}$.




\subsection{Excited State Captures}
\label{excited}
As seen in Fig.~\ref{decompfig}, captures to the second excited
state are readily identified by the presence of a 6.13~MeV
prompt event---these events are well separated from the prompt
neutron continuum and there is little contamination of the
sample. These events constitute approximately 1\% of the total.

Captures to the first excited state can be identified by the
1022~keV prompt signal from annihilation of the positron. These
events are harder to separate from the fast neutron continuum,
but a useful peak is present with sufficient statistics. This
feature of the source spectrum differs from that expected from
the extant background in KamLAND because the kinetic energy of
the electron positron pair is fully contained inside the source
capsule. Roughly 5\% of the total rate is associated with these
events.

\subsection{Tagged Slow Neutrons}
\label{slown}
The presence of a 6.13 MeV gamma in the final state of the alpha
capture reaction is a clean and experimentally accessible
indication of an excited state capture and is associated with an
initial neutron energy below 0.6 MeV (by comparison both
\isotope[252]{Cf} and \isotope[241]{Am}--\isotope{Be} sources
produce neutrons with average energy well over 1~MeV). We suggest
that the $\isotope[13]{C}(\alpha,\isotope{n})\isotope[16]{O}$
reaction is an effective source of tagged slow neutrons and may
be useful in the calibration of future reactor neutrino
experiments e.g. Double Chooz or Daya
Bay.\cite{Ardellier:2006mn,Guo:2007ug}.

When constructed as a tagged neutron source, an alpha progenitor
with a longer half-life than \isotope[210]{Po} is
preferable. Isotopes with significant fission rates should be
avoided as they introduce unwanted neutron
background. Progenitors with significantly higher alpha energies
can achieve much higher branching fractions to the second excited
state, but also have higher endpoint energies for neutrons
emitted from this state--possibly negating the advantage of
tagging a `slow' neutron.  \isotope[210]Pb is a viable choice,
but may be difficult to obtain in sufficient
quantity. \isotope[241]{Am} is easily obtainable and long-lived
and thus appears to be a useful progenitor, though care must be
taken to shield the detector from 59~keV gamma line. Another
possibility is \isotope[238]{Pu}.

\section{Summary}
\label{conclusions}
We have constructed a calibration source based on the reaction
\isotope[13]{C}($\alpha$,\isotope{n})\isotope[16]{O} using
\isotope[210]{Po} as the alpha progenitor and successfully
deployed it in KamLAND. Both the neutron rate and the 6.13~MeV
gamma rate were consistent with our expectations and were
sufficient to accumulate useful statistics in a reasonable time.

Data from the source are being employed to reduce the systematic
uncertainty associated with the presence of radon decay daughters
in the detector. Efforts to use these data to improve KamLAND
event reconstruction and shed additional light on the properties
of the KamLAND liquid scintillator---neutron quenching and energy
scale---are underway.


\section*{Acknowlegments}
This work was done as part of our calibration efforts for KamLAND
and could not have proceeded without the support of collaboration
as a whole. We would particularly like to acknowledge the
assistance of Kazumi Tolich in writing and debugging trigger
scripts. We are also indebted to Kevin Lesko and Brian Fujikawa
for bringing the possibility of building an
$\isotope[13]{C}(\alpha,\isotope{n})\isotope[16]{O}$ source for
KamLAND to our attention. The \isotope[13]{C} powder used in
constructing the source was supplied by Kevin Lesko. Andreas
Piepke aided us in developing the source certification
procedure. This work was funded by the US Department of Energy.


\begin{table}[h]
  \centering
  \caption{Summary of relevant data for states of
    \isotope[16]{O}. Threshold represents the minimum alpha
    kinetic energy needed to excite the state in
    $\isotope[13]{C}(\alpha,\isotope{n})\isotope[16]{O}$.}
  \begin{tabular}{lr@{.}lcc} 
    State & \multicolumn{2}{r}{Threshold (MeV)} & JP & Decay mode 
    \\
    \hline
    Ground & \multicolumn{2}{l}{~~--} & 0+ & -- \\
    1st excited & 5&014 & 0+ & $\mbox{e}^- + \mbox{e}^+$ \\
    2nd excited & 5&119 & 3- & $\gamma$ \\
    3rd excited & 6&148 & 2+ &  \\
  \end{tabular}
  \label{o16states} 
\end{table}

\end{document}